\documentclass[twocolumn,showpacs,preprintnumbers,amsmath,amssymb,prl,superscriptaddress]{revtex4-1}
\usepackage{bbm}
\usepackage{mathrsfs}
\usepackage{graphicx}
\usepackage{dcolumn}
\usepackage{bm}
\usepackage{amsmath}
\usepackage{amsfonts}
\usepackage{color}

\begin{document}

\title{Dynamical Axion Field in a Magnetic Topological Insulator Superlattice}
\author{Jing Wang}
\affiliation{Department of Physics, McCullough Building, Stanford University, Stanford, California 94305-4045, USA}
\affiliation{Stanford Institute for Materials and Energy Sciences, SLAC National Accelerator Laboratory, Menlo Park, California 94025, USA}
\author{Biao Lian}
\affiliation{Department of Physics, McCullough Building, Stanford University, Stanford, California 94305-4045, USA}
\author{Shou-Cheng Zhang}
\affiliation{Department of Physics, McCullough Building, Stanford University, Stanford, California 94305-4045, USA}
\affiliation{Stanford Institute for Materials and Energy Sciences, SLAC National Accelerator Laboratory, Menlo Park, California 94025, USA}

\begin{abstract}
We propose that the dynamical axion field can be realized in a magnetic topological insulator superlattice or a topological paramagnetic insulator.
The magnetic fluctuations of these systems produce a pseudoscalar field which has an axionic coupling to the electromagnetic field, and thus it gives a condensed-matter realization of the axion electrodynamics. Compared to the previously proposed dynamical axion materials where a long range antiferromagnetic order is required, the systems proposed here have the advantage that only a uniform magnetization or a paramagnetic state is needed for the dynamic axion. We further propose several experiments to detect such a dynamical axion field.
\end{abstract}

\date{\today}


\maketitle


The search for topological quantum phenomena has attracted considerable interest in condensed matter physics.
Topological phenomena are determined by some topological structures in physical systems and are thus usually universal and robust
against perturbations~\cite{thouless1998}. The recent discovery of the time-reversal ($\mathcal{T}$) invariant
(TRI) topological insulator (TI) brings the opportunity to realize a large family of new topological phenomena~\cite{hasan2010,qi2011}. The electromagnetic response of three-dimensional (3D) insulators is described by the Maxwell action $\mathcal{S}_{\text{M}}=(1/8\pi)\int d^3xdt(\epsilon\mathbf{E}^2-\mathbf{B}^2/\mu)$, together with a topological $\theta$ term $\mathcal{S}_{\theta}=(\theta/2\pi)(\alpha/2\pi)\int d^3xdt\mathbf{E}\cdot\mathbf{B}$~\cite{qi2008}. Here, $\mathbf{E}$ and $\mathbf{B}$ are
the conventional electromagnetic fields inside the insulator, $\epsilon$ and $\mu$ are material-dependent dielectric constant and magnetic permeability, $\alpha=e^2/\hbar c$ is the fine structure constant, $e$ is the charge of an electron,
and $\theta$ is the dimensionless pseudoscalar parameter describing the insulator, which refers
to the axion field in high energy physics~\cite{peccei1977,wilczek1987}. Physically $\theta$ depends on the band structure of the
insulator and has an explicit microscopic expression of the momentum space Chern-Simons form~\cite{qi2008}:
\begin{equation}\label{theta}
\theta=\frac{1}{4\pi}\int
d^3k\epsilon^{ijk}\mathrm{Tr}\left[\mathcal{A}_i\partial_j\mathcal{A}_k+i\frac{2}{3}\mathcal{A}_i\mathcal{A}_j\mathcal{A}_k\right],
\end{equation}
where $\partial_j=\partial/\partial k_j$, $\mathcal{A}_i^{\mu\varrho}(\mathbf{k})=-i\langle u^{\mu}_{\mathbf{k}}|\partial_i|u^{\varrho}_\mathbf{k}\rangle$ is the momentum space non-abelian gauge field, with $|u^{\mu}_{\mathbf{k}}\rangle$ and $|u^{\varrho}_{\mathbf{k}}\rangle$ referring to the periodic part of the Bloch function of the occupied bands. All physical quantities in the bulk depend on $\theta$ only modulo $2\pi$. $\mathcal{S}_{\theta}$ generally breaks the parity $\mathcal{P}$ and $\mathcal{T}$ symmetry except for two TRI points $\theta=0$ and $\theta=\pi$, which describe trivial insulator and TI, respectively~\cite{qi2008}. The term $\mathcal{S}_{\theta}$ with a universal value of $\theta=\pi$ in TIs gives rise to new physical effects such as image magnetic monopole~\cite{qi2009b}, quantized Kerr effect~\cite{maciejko2010,tse2010}, and quantized topological magnetoelectric effect~\cite{qi2008,nomura2011,wang2015b,morimoto2015}.

The axion field $\theta$ is \emph{static} in a TRI TI. However, as is first suggested in Ref.~\cite{li2010}, when a long range antiferromagnetic (AFM) order is established in a TI, $\theta$ can deviate from $\pi$ due to $\mathcal{T}$ symmetry breaking and becomes a \emph{dynamical} field associated with the magnetic fluctuations. The resulting system is a new state of quantum matter which realizes the axion electrodynamics in condensed matter physics. The axionic excitations in such an unconventional AFM insulator can
lead to novel effects such as the axionic polariton~\cite{li2010}. There have been great efforts devoted to searching for such dynamical axion state of matter~\cite{wang2011a,ooguri2012,wan2012,kim2012,shiozaki2014,goswami2014}. However, such AFM materials are still lacking.

\begin{figure}[t]
\begin{center}
\includegraphics[width=3.3in]{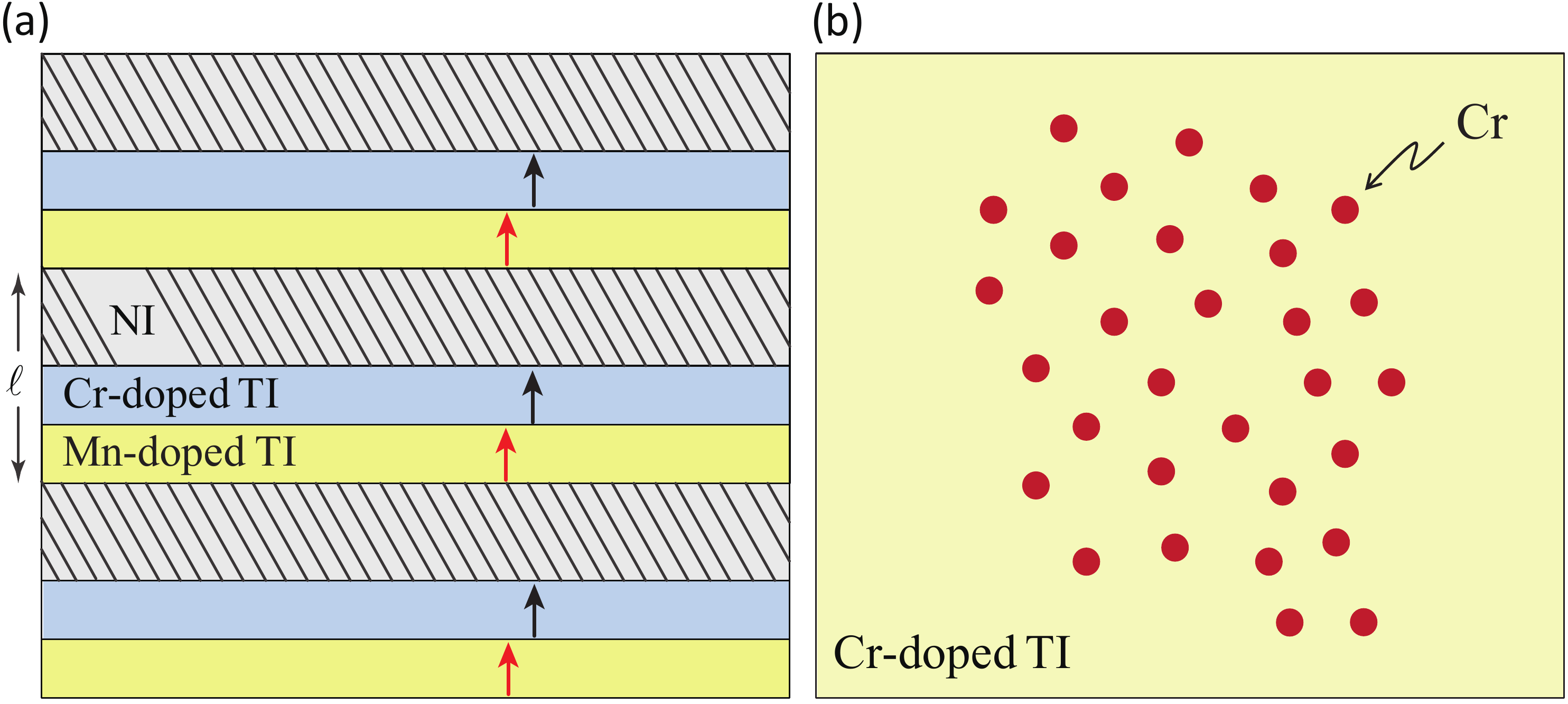}
\end{center}
\caption{(color online). Schematic drawing of the proposed systems to realize the dynamical axion field. (a) The magnetic TI superlattice structure. The upper and lower halves of TI films are doped with Cr and Mn, respectively. The arrows in TI layers indicate the magnetization direction. In each crystalline unit cell, the $z$-direction thickness of Mn-doped TI, Cr-doped TI, and NI layers are $\ell_m$, $\ell_c$, and $\ell_n$. (b) Topological PM insulator: 3D TI materials close to topological QCP doped with Cr, where a PM state is realized.}
\label{fig1}
\end{figure}

In this paper, we propose a much simpler route to realize the dynamic axion in a magnetic TI superlattice.
In particular, we clarify that the realization of a dynamical axion field does \emph{not} necessarily require an AFM order and a TI parent material, but what is important is a proper coupling between the electrons and magnetic fluctuations~\cite{li2010,wang2011a,ooguri2012}. The magnetic TI superlattice we adopt consists of alternating layers of a parent magnetic TI and a spacer normal insulator (NI), as shown in Fig.~\ref{fig1}(a). The magnetic TI layer is doped with Cr and Mn on top and bottom halves of TI film, respectively. We show that the phase diagram of this system contains a dynamic axion phase when uniform magnetization is achieved, where $\mathcal{T}$ symmetry is broken and the static value $\theta_0\neq0,\pi$. We also propose to realize the dynamic axion in a topological paramagnetic (PM) insulator, where $\mathcal{T}$ symmetry is present and $\theta_0=0$ or $\pi$. The PM fluctuations can couple to electrons, which induces a dynamical axion field. Such a PM insulator can be achieved by doping magnetic elements into TI materials close to the topological quantum critical point (QCP) [shown in Fig.~\ref{fig1}(b)]. Finally, we propose several experiments to detect this dynamical axion field.

The first system we propose is a magnetic TI superlattice as described above.
Recent experiments have shown that the thickness and magnetic doping concentration of thin film TIs can be well controlled through layer-by-layer growth via molecular beam epitaxy~\cite{zhang2010}, therefore, such a superlattice is quite realistic to be fabricated. The magnetic ions will have a local exchange coupling with the band electrons described by $J_{\nu}\sum_{\mathbf{x}_i}\mathbf{S}_{\nu}(\mathbf{x}_i)\cdot\mathbf{s}$, where $\mathbf{S}_{\nu}(\mathbf{x}_i)$ denotes the magnetic impurity spin at the position $\mathbf{x}_i$, $\nu=c,m$ denotes the Cr and Mn, respectively, and $\mathbf{s}=\boldsymbol{\sigma}/2$ is the local electron spin. The main advantage of such a superlattice is the two types of magnetic ions have opposite signs of exchange coupling parameters in TI materials, namely, $J_{m}<0$ and $J_{c}>0$.  This is experimentally verified by opposite signs of the anomalous Hall conductance in the insulating regime of Mn-doped~\cite{checkelsky2012} and Cr-doped~\cite{chang2013b} Bi$_2$Te$_3$ family materials. Therefore a uniform magnetization in the superlattice will induce opposite exchange fields in the upper and lower halves of a TI layer. The Hamiltonian describing the superlattice can be written as
\begin{eqnarray}\label{model}
\mathcal{H} &=& \sum\limits_{\mathbf{k}_\parallel,i,j}\Big[v_F\tau^z\left(\hat{\mathbf{z}}\times\boldsymbol{\sigma}\right)\cdot\mathbf{k}_{\parallel}\delta_{i,j}
+m_a\tau^z\sigma^z\delta_{i,j}+m_b\sigma^z\delta_{i,j}
\nonumber
\\
&&+t_s\tau^x\delta_{i,j}+\frac{t_n}{2}\tau^+\delta_{i+1,j}+\frac{t_n}{2}\tau^-\delta_{i-1,j}\Big]
c^{\dag}_{\mathbf{k}_{\parallel i}}c_{\mathbf{k}_{\parallel j}}.
\end{eqnarray}
where $i$ and $j$ label distinct magnetic TI layers, $\sigma^\beta$  and $\tau^\beta$ ($\beta=x,y,z$) are Pauli matrices acting on the spin and the top/bottom surface of the parent TI layer, respectively. The first term in the Hamiltonian describes the top and bottom surface states of a parent TI layer, where a single 2D Dirac node is considered for Bi$_2$Te$_3$ family materials~\cite{zhang2009}. $v_F$ is the Fermi velocity. $\mathbf{k}_{\parallel}=(k_x,k_y)$ is the in-plane momentum. The second and third terms describe the Zeeman-type spin splitting for surface states induced by the ferromagnetic (FM) exchange couplings $\Delta_{c}$ of Cr and $\Delta_m$ of Mn along $z$ axis, where $m_a=(\Delta_c-\Delta_m)/2$ is the staggered Zeeman field and $m_b=(\Delta_c+\Delta_m)/2$ is the uniform Zeeman field~\cite{wang2014a}. In the mean field (MF) approximation, the exchange field of Mn and Cr are given by $\Delta_{\nu}=y_{\nu}J_{\nu}\langle S_\nu^z\rangle/2$. Here $y$ is the doping concentration of magnetic ion, $\langle S^z\rangle$ is the MF expectation value of the ion spin in the $z$ direction, $\nu=c,m$. The thickness dependent parameters $t_s$ and $t_n$ describe the tunneling between the top and bottom surface states within the same ($t_s$) or neighboring ($t_n$) TI layer. For simplicity, we assume $J_c=-J_m=J>0$.

\begin{figure}[t]
\begin{center}
\includegraphics[width=3.3in]{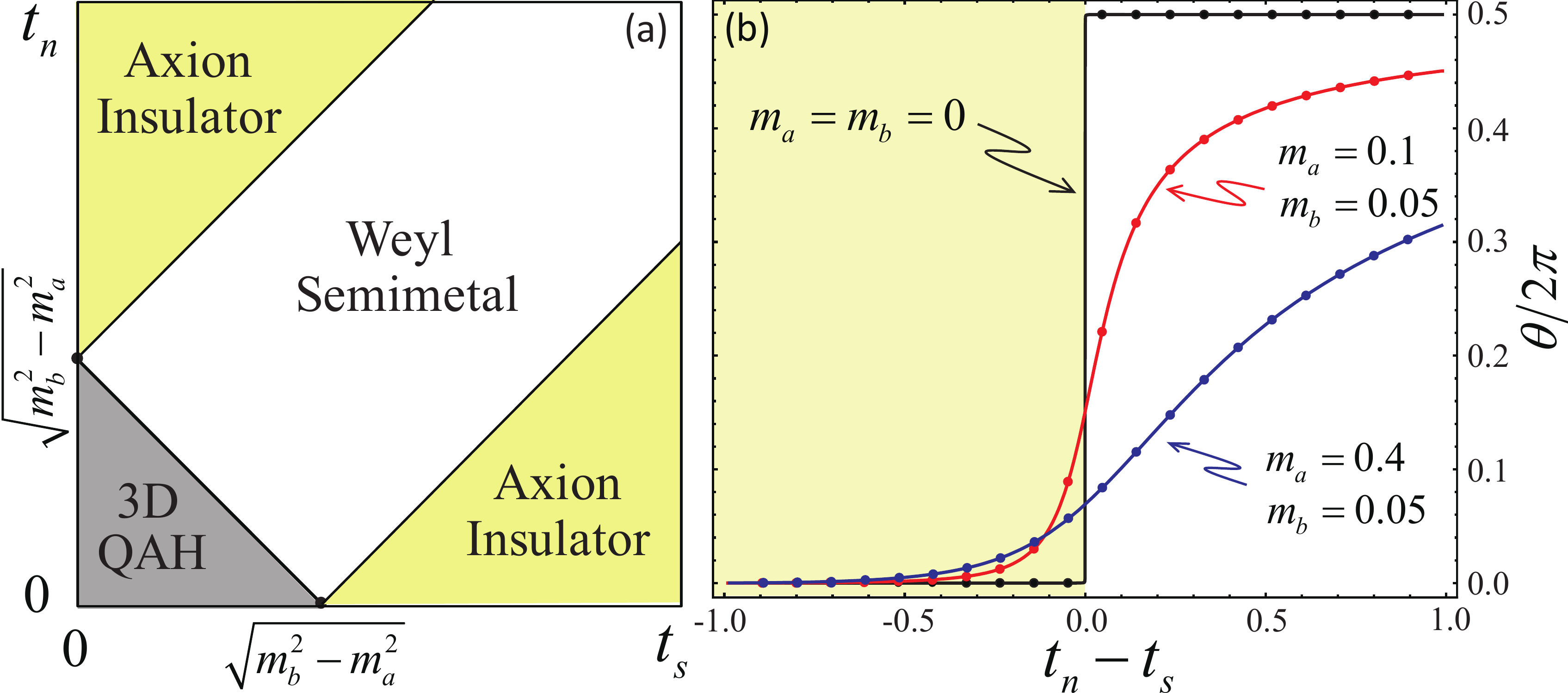}
\end{center}
\caption{(color online). (a) Phase diagram of the proposed magnetic TI superlattice with two variables: $t_s$ and $t_n$. An AI phase emerges with nonzero $\theta$. When $m_a>m_b$, the phase diagram will be AI phase only (not shown). (b) Typical value of $\theta$ as a function of $t_n-t_s$ in the AI phase, where the parameter is setted as $t_n+t_s=1$.}
\label{fig2}
\end{figure}

First, we examine the phase diagram of the system. The momentum space Hamiltonian now is
\begin{eqnarray}\label{model_k}
\mathcal{H}(\mathbf{k}) &=& \sum\limits_{a=1}^5 d_a(\mathbf{k})\Gamma^a+m_b\Gamma^{12},
\end{eqnarray}
where $d_{1,2,3,4,5}(\mathbf{k})=(v_Fk_y,-v_Fk_x,-t_n\sin(k_z\ell),t_s+t_n\cos(k_z\ell),m_a)$, and the Dirac $\Gamma$ matrices $\Gamma^{1,2,3,4,5}=(\tau^z\sigma^x,\tau^z\sigma^y,\tau^y,\tau^x,\tau^z\sigma^z)$, $\Gamma^{12}=[\Gamma^1,\Gamma^2]/2i=\sigma^z$. The $\mathcal{T}$ and $\mathcal{P}$ transformation are defined as $\mathcal{T}=i\sigma^y\mathcal{K}$ (with
$\mathcal{K}$ being the complex conjugation operator) and $\mathcal{P}=\tau^x$, respectively. The band dispersion is given by
\begin{equation}
\varepsilon_{\mathbf{k}\pm}^2 = v_F^2(k_x^2+k_y^2)+\left[m_b\pm\sqrt{m_a^2+t^2(k_z)}\right]^2,
\end{equation}
where $t(k_z)=\sqrt{t_s^2+t_n^2+2t_st_n\cos(k_z\ell)}$ and $\ell$ is the superlattice period along the growth $z$ direction with $\ell=\ell_m+\ell_c+\ell_n$. In the absence of exchange field, i.e., $m_a=m_b=0$, the system is fully gapped when $\left|t_s\right|\neq\left|t_n\right|$, while it has a gapless Dirac node when $t_s/t_n=\pm1$~\cite{burkov2011}. For convenience we assume here $t_s/t_n\geq0$. When $t_s/t_n=1$, the Dirac point is located at $k_x=k_y=0$, $k_z=\pi/\ell$. Such a critical Dirac point opens a gap when $t_s/t_n$ deviates from unity, resulting in a 3D TI ($t_n>t_s$) or NI ($t_n<t_s$). Since both $\mathcal{P}$ and $\mathcal{T}$ symmetries are respected, the axion field $\theta$ given by Eq.~(\ref{theta}) is either $0$ or $\pi$ in this case, as shown in Fig.~\ref{fig2}(b). In the case of $m_{a,b}\neq0$, the band structure has two nondegenerate Weyl nodes when $t_{c1}^2\equiv\left(t_s-t_n\right)^2<m_b^2-m_a^2<\left(t_s+t_n\right)^2\equiv t^2_{c2}$, located on the $k_z$ axis at $k_z=\pi/\ell\pm k_0$ where $k_0\ell=\arccos[(m_b^2-m_a^2-t_s^2-t_n^2)/2t_st_n]$. Such a Weyl semimetal phase occurs in a finite region in the phase diagram as shown in Fig.~\ref{fig2}(a). When $m_b^2-m_a^2>t^2_{c2}$, the system is a 3D quantum anomalous Hall (QAH) insulator characterized by a quantized Hall conductivity $e^2/h$ per magnetic TI layer. Interesting physics happens when $m_b^2-m_a^2<t_{c1}^2$. The system is fully gapped, however, as we will show below, it is not a simple NI but an axionic insulator (AI) with $\theta\neq0,\pi$. Furthermore, the FM fluctuations in the AI lead to a dynamical axion field.

Since $\theta$ is odd under $\mathcal{T}$ and $\mathcal{P}$, only $\mathcal{T}$- and $\mathcal{P}$-breaking perturbations can induce a change of $\theta$. The term $m_a\Gamma^5$ breaks both $\mathcal{T}$ and $\mathcal{P}$, which varies the value of $\theta$ to the linear order. $m_b\Gamma^{12}$ breaks $\mathcal{T}$ but respects $\mathcal{P}$, therefore it does not affect $\theta$ to the leading order. To compute the axion field $\theta$, a lattice regularization is necessary. Explicitly, the value of $\theta$ in this model can be calculated as~\cite{li2010,wang2011a},
\begin{equation}\label{theta_Dirac}
\theta=\frac{1}{4\pi}\int
d^3k\frac{2|d|+d_4}{(|d|+d_4)^2|d|^3}\epsilon^{ijkl}d_i\partial_xd_j\partial_yd_k\partial_zd_l,
\end{equation}
where $|d|=(\sum_{a=1}^5d_a^2)^{1/2}$, and the repeated index $i,j,k,l$ take values from $1,2,3,5$ and indicates summation. Typical $\theta$ values in AI phase is calculated in Fig.~\ref{fig2}(b). As expected, $\theta$ deviates gradually from $0$ or $\pi$ as $m_a$ increases away from $0$. When $t_n/t_s\gg1$, the $\theta$ value tends to $\pi$; for $t_n/t_s\rightarrow0$, $\theta$ converges quickly towards zero. Therefore, $\theta$ can be tuned by the layer thickness. Note that $\theta$ is only well defined in the insulating regime when $m_b^2-m_a^2<t_{c1}^2$. For $m_a>m_b$, this condition is always satisfied and the whole phase diagram in Fig.~\ref{fig2}(a) will be occupied by the AI phase only. Physically, $m_a>m_b$ means the magnetic moments in Mn and Cr are polarized in the same direction. Different from the previous proposals~\cite{li2010,wang2011a}, such nonquantized $\theta$ is coupled to the FM order parameter instead of AFM order, due to opposite signs of $J_c$ and $J_m$.

To realize the AI phase, the system should have an appropriate magnetic ordering. If $m_a>m_b$, the system will have a FM long range order along $z$ axis and become an AI. Conversely if $m_a<m_b$, the system will have an AFM long range order, where the spins of Mn and Cr in each parent magnetic TI layer will point along the $+z$ and $-z$ direction, respectively. In this case, $m_b^2-m_a^2<t_{c1}^2$ must be satisfied to realize the AI, which may be fulfilled by adjusting the doping concentration $y_{c,m}$ and tuning the layer thicknesses $\ell_{c,m,n}$. The magnetic properties of this system are determined by the effective interaction between neighbouring magnetic impurity spins $\mathcal{J}_{\text{eff}}^{\rho\nu}[S^z_{\rho}(\mathbf{x}_i) S^z_{\nu}(\mathbf{x}_j)+ \gamma\mathbf{S}^\parallel_{\rho}(\mathbf{x}_i)\cdot\mathbf{S}^\parallel_{\nu}(\mathbf{x}_j)]$, where $\mathbf{S}^\parallel_{\rho}$ denotes the in-plane impurity spin, and $\rho,\nu=c,m$ labels the ion type. Such effective spin interactions are mediated by the band electrons of TIs~\cite{liu2009,yu2010,wang2015}. The interactions between the same types of magnetic ions have been shown to be FM with an easy axis $z$, which indicates $\mathcal{J}_{\text{eff}}^{mm}<0$ for Mn-doped TI film~\cite{checkelsky2012}, $\mathcal{J}_{\text{eff}}^{cc}<0$ for Cr-doped TI film~\cite{chang2013b} and $|\gamma|<1$.
The sign of $\mathcal{J}_{\text{eff}}^{mc}$ is determined by the Ruderman-Kittel-Kasuya-Yoshida (RKKY) type interaction~\cite{dietl2014} along the $z$ direction. $\mathcal{J}_{\text{eff}}^{mc}(\ell_z)\propto-J_mJ_c\int_{-\pi/a_3}^{\pi/a_3} dq_z\chi_{zz}(q_z)e^{iq_z\ell_z}$, where $a_3=1$~nm is the $z$ direction lattice constant of parent TI material, i.e., thickness of a quintuple layer (QL). $\chi_{zz}(q_z)$ is the $z$ direction magnetic susceptibility of TI with momentum $\mathbf{q}=(0,0,q_z)$ obtained by Kubo formula~\cite{dietl2014}. $\ell_z$ is the vertical distance between Mn and Cr, which we set to their mean distance in a magnetic TI layer as $\ell_z=(\ell_m+\ell_c)/2$. The calculated sign of $\mathcal{J}_{\text{eff}}^{mc}(\ell_z)$ oscillates as a function of $\ell_z$~\cite{li2015}, is listed in Table~\ref{table1}. The sign of $\mathcal{J}_{\text{eff}}^{mc}$ is opposite to that in Ref.~\cite{li2015} since $J_mJ_c<0$. We note that the exact sign of the interlayer coupling has not been settled yet
by experiments. According to Table~\ref{table1}, $\mathcal{J}_{\text{eff}}^{mc}<0$ for $\ell_m=\ell_c=1,3$~QL, possibly leading to a FM ground state, and the system becomes an AI. For $\ell_m=\ell_c=2$~QL, the system may develop an AFM order, yet one can still reach an AI state by tuning $t_n$.

\begin{table}[t]
\caption{The parameters of $t_s$, sign of mean $\mathcal{J}^{mc}_{\mathrm{eff}}$ and possible magnetic order along $z$ direction in Cr-doped and Mn-doped (Bi$_{0.1}$Sb$_{0.9}$)$_2$Te$_3$ superlattice with different thickness. Here we set $\ell_m=\ell_c$, and assume $t_n\neq0$.}
\begin{center}\label{table1}
\renewcommand{\arraystretch}{1.2}
\begin{tabular*}{3.3in}
{@{\extracolsep{\fill}}ccccc}
\hline
\hline
$\ell_m+\ell_c$ & $t_s$ (eV) & $\text{sgn}\left(\mathcal{J}^{mc}_{\mathrm{eff}}\right)$ & possible order & AI
\\
\hline
2 QL & $0.116$ & $-$ & FM & Yes
\\
4 QL & $0.029$ & $+$ & AFM & ?
\\
6 QL & $0.004$ & $-$ & FM & Yes
\\
\hline
\hline
\end{tabular*}
\end{center}
\end{table}


Next we show the axion $\theta$ becomes a dynamical field $\theta=\theta_0+\delta\theta(\mathbf{x},t)$ in the presence of the FM fluctuations. The magnetic fluctuations in the TI originate from the quantum nature of spin interactions (for $\gamma\neq0$) and the thermal fluctuations. For convenience we define the magnetization per unit volume of Cr and Mn as $\mathbf{M}_{\nu}=g_L\mu_By_\nu\langle\mathbf{S}_\nu\rangle/a^3$ with $\nu=c,m$. Here $g_L$ is the Land\'{e} factor, $\mu_B$ is the Bohr magneton, $a$ is the average lattice constant of TI. They can be regrouped into the FM and AFM magnetization as $\mathbf{M}_{\pm}=(\mathbf{M}_c\pm\mathbf{M}_m)/2$. In the below, we assume $y_c=y_m=y$. The fluctuation of $\mathbf{M}_\pm$ can be generally written as $\mathbf{M}_{\pm}=(M_0^{\pm}+\delta M^\pm_z(\mathbf{x},t))\hat{\mathbf{z}}+\delta M^\pm_x(\mathbf{x},t)\hat{\mathbf{x}}+\delta M^\pm_y(\mathbf{x},t)\hat{\mathbf{y}}$. To the linear order, it can be deduced from Eq.~(\ref{theta_Dirac}) that the axion field $\theta$ is only coupled to $d_5=m_a=(\Delta_c-\Delta_m)/2\propto M^+_z$. Therefore, only the FM fluctuations along $z$ axis $\delta M^+_z$ are relevant. The corresponding effective Lagrangian is $\mathcal{L}_M=\mathcal{K}_M[(\partial_t\delta M_z^+)^2-(v_i\partial_i\delta M_z^+)^2-m_s^2(\delta M_z^+)^2]$, where $\mathcal{K}_M$, $v_i$ and $m_s$ are the stiffness, velocity and mass of the spin-wave mode $\delta M^+_z$. The fluctuation of $\theta$ is now given by $\delta\theta(\mathbf{x},t)=\delta m_a(\mathbf{x},t)/g=(Ja^3/4g_L\mu_Bg)\delta M_z^+(\mathbf{x},t)$, where the coefficient $g$ can be determined from Eq.~(\ref{theta_Dirac}). The effective Lagrangian density describing the axion coupled electromagnetic response is then given by
\begin{eqnarray}\label{axion_L}
\mathcal{L} &=& \mathcal{L}_{\text{Maxwell}}+\mathcal{L}_{\theta}+\mathcal{L}_{\text{axion}}
\nonumber
\\
&=&\frac{1}{8\pi}\left(\epsilon\mathbf{E}^2-\frac{1}{\mu}\mathbf{B}^2\right)+\frac{\alpha}{4\pi^2}\left(\theta_0+\delta\theta\right)\mathbf{E}\cdot\mathbf{B}
\nonumber
\\
&&+g^2\mathcal{K}_A\left[(\partial_t\delta\theta)^2-(v_i\partial_i\delta\theta)^2-m_s^2\delta\theta^2\right],
\end{eqnarray}
where the three terms describe the conventional Maxwell action, the topological coupling between the axion and the electromagnetic
field, and the dynamics of the massive axion. $\mathcal{K}_A=\mathcal{K}_M(4\mu_Bg_L/Ja^3)^2$. The axion mass at temperature $T$ is $m_s\sim\left|k_BT-\mathcal{J}_F\right|/\hbar$, where $\mathcal{J}_F=|\mathcal{J}_{\text{eff}}^{cc}+\mathcal{J}_{\text{eff}}^{mm}-2\mathcal{J}_{\text{eff}}^{mc}|/2$ is of the same order as the Curie temperature and decays exponentially with the mean distance between magnetic ions $\xi=y^{-1/3}a$. The coefficient $\mathcal{K}_M\sim\xi^3\hbar/g_L^2\mu_B^2m_s$, while the velocity $v_i\sim \xi m_s$. For an estimation, in a typical magnetic TI system, $m_s\sim k_BT\sim\mathcal{J}_F\sim1$~meV, the bulk gap is $\left|t_s-t_n\right|\sim0.1$~eV and the Zeeman field is $m_a\sim0.06$~eV. Therefore, $m_s\ll|m_a|<|t_s-t_n|$, justifying the above low-energy description of the system.

The coupling between the dynamic axion field $\theta$ and the electromagnetic field gives rise to a number of novel topological phenomena, which can be used in experiments as a unique signature of dynamic $\theta$. For instance, it leads to the formation of axion polariton, which becomes gapped in the presence of a background magnetic field. It also leads to the double frequency response on the cantilever torque magnetometry~\cite{li2010}. Here we mention another interesting phenomena proposed in Ref.~\cite{ooguri2012}, that the massive axion in $\mathcal{L}$ exhibits an instability in the presence of an external electric field $E_0$. Such an instability will lead to a complete screening of electric field above a critical value $E_{\text{crit}}$. In other words, when $\epsilon_0E_0<\epsilon E_{\text{crit}}$, the field inside the system is $E=\epsilon_0E_0/\epsilon$ and $B=0$; when $\epsilon_0E_0>\epsilon E_{\text{crit}}$, one gets $E=E_{\text{crit}}$ and $B=\pm\sqrt{\mu E_{\text{crit}}(\epsilon_0E_0-\epsilon E_{\text{crit}})}$. Here $\epsilon_0$ is the dielectric constant outside the system, and $E_{\text{crit}}=(m_s/\alpha)\sqrt{8\pi^3g^2\mathcal{K}_A/\mu}$. For $\theta_0=0$, a second-order phase transition happens at $\epsilon_0E_0=\epsilon E_{\text{crit}}$, while for $\theta_0\neq0$, the phase transition becomes a crossover. More details on these experimental proposals are presented in the Supplemental Material~\cite{supplement}.
The relative permittivity, axion mass, and axion coupling of the magnetic TI system are estimated to be $\epsilon\sim100$, $m_s\sim1$~meV, $g\sim0.08$~eV, $J\sim1.5$~eV, $y\sim0.1$, and $a=0.5$~nm. This gives $E_{\text{crit}}\sim(g/J)\sqrt{m_s/\epsilon ya^3}=2\times10^6$~V/m, which is much smaller than the breakdown field of the typical semiconductors and in the range accessible by experiments. The critical field $E_{\text{crit}}\propto\sqrt{m_s/y}$ could be reduced by adjusting the doping concentration $y$ of the system. For extremely low temperatures, $m_s\sim\mathcal{J}_F\propto e^{-\lambda y^{-1/3}}$, and $E_{\text{crit}}$ becomes smaller as $y$ decreases. For relatively high temperatures when $m_s\sim k_BT$ is independent of $y$, $E_{\text{crit}}$ will be reduced as $y$ increases.

In the above discussion, we show that to realize the dynamic axion, it is not essential to start from a nontrivial TI or a magnetic order. In fact, in such a TI system, the effect of the dynamical axion may be suppressed in the bulk since the electromagnetic field mainly couples to the surface states~\cite{li2010}. Instead, a topologically trivial insulator with magnetic fluctuations properly coupled to the electrons is also able to produce dynamic axions, and the low-energy physics is dominated by the bulk. This motivates us to propose the second dynamic axion system which is a PM insulator. Such a system can be realized by doping magnetic elements such as Cr into 3D TI materials to the vicinity of the topological QCP, for example, Bi$_2$(Se$_{x}$Te$_{1-x}$)$_3$ with $x\geq0.66$~\cite{zhang2013}. The system is topologically trivial and exhibits a PM response at low temperature, which is caused by the reduced effective spin-orbit coupling strength of Cr$_y$Bi$_{2-y}$ resulting from the Cr substitution of Bi. The Hamiltonian of the system is the Dirac model $\mathcal{H}_{b}=\sum_{a=1}^5\widetilde{d}_a(\mathbf{k})\widetilde{\Gamma}^a$ as in Ref.~\cite{li2010}, $\widetilde{d}_a(\mathbf{k})=(\sin k_x,\sin k_y,\sin k_z, m_4(\mathbf{k}), m_5)$,  $\widetilde{\Gamma}_a=(\tau_x\sigma_x,\tau_x\sigma_y,\tau_y,\tau_z,\tau_x\sigma_z)$, $\tau^i$ refers to orbit index. $m_4$ is topologically trivial mass, while $m_5=0$ on average, leading to a mean value $\theta_0=0$.
The AFM fluctuation $\delta M_z^-$ of Cr spins inside a unit cell will induce a fluctuation $\delta m_5$, leading to a dynamical axion field $\delta\theta=\delta m_5/g$~\cite{supplement}. The advantage of such a system is that it is close to the PM to FM transition~\cite{zhang2013},
therefore the magnetic fluctuation is strong and the axion mass $m_s$ is small. To distinguish with the previous proposals, this material may be called topological PM insulator which is a TRI AI with a dynamic axion field.

In summary, we show that the dynamical axion field can be realized in a magnetic TI superlattice. We emphasize that each magnetic TI layer does not need to exhibit QAH effect, but only a uniform magnetization is necessary. We hope
the theoretical work here could aid the search for the axionic state of matter in real materials.

\begin{acknowledgments}
This work is supported by the US Department of Energy, Office of Basic Energy Sciences, Division of Materials Sciences and Engineering, under Contract No.~DE-AC02-76SF00515 and in part by the NSF under grant No.~DMR-1305677.
\end{acknowledgments}

\end{document}